\RequirePackage{ifpdf}
\ifpdf 
\documentclass[pdftex]{sigma}
\else
\documentclass{sigma}
\fi

\def\p{\partial}

\newcommand{\C}{\mathbb{C}}

\newcommand{\kt}{\rangle}
\newcommand{\br}{\langle}
\newcommand{\bbr}{\br \br}
\newcommand{\kkt}{\kt \kt}
\newcommand{\pbr}{\prec}
\newcommand{\pkt}{\succ}

\begin{document}

\allowdisplaybreaks

\renewcommand{\thefootnote}{$\star$}

\renewcommand{\PaperNumber}{001}

\FirstPageHeading

\ShortArticleName{Three-Hilbert-Space Formulation of Quantum
Mechanics}

\ArticleName{Three-Hilbert-Space Formulation \\ of Quantum
Mechanics\footnote{This paper is a contribution to the Proceedings of the VIIth Workshop ``Quantum Physics with Non-Hermitian Operators''
     (June 29 -- July 11, 2008, Benasque, Spain). The full collection
is available at
\href{http://www.emis.de/journals/SIGMA/PHHQP2008.html}{http://www.emis.de/journals/SIGMA/PHHQP2008.html}}}

\Author{Miloslav ZNOJIL}

\AuthorNameForHeading{M. Znojil}

\Address{Nuclear Physics Institute ASCR,
250 68 \v{R}e\v{z}, Czech Republic}
\Email{\href{mailto:znojil@ujf.cas.cz}{znojil@ujf.cas.cz}}
\URLaddress{\url{http://gemma.ujf.cas.cz/~znojil/}}

\ArticleDates{Received October 29, 2008, in f\/inal form December 31,
2008; Published online January 06, 2009}

\Abstract{In paper [Znojil M., \textit{Phys. Rev. D} \textbf{78} (2008), 085003, 5~pages, \href{http://arxiv.org/abs/0809.2874}{arXiv:0809.2874}] the
two-Hilbert-space (2HS, a.k.a.~cryptohermitian) formulation of
Quantum Mechanics has been revisited. In the present continuation of
this study (with the spaces in question denoted as ${\cal
H}^{\rm (auxiliary)}$ and ${\cal H}^{\rm (standard)}$) we spot a weak point
of the 2HS formalism which lies in the double role played by ${\cal
H}^{\rm (auxiliary)}$. As long as this conf\/luence of roles may (and
did!) lead to confusion in the literature, we propose an amended,
three-Hilbert-space (3HS) reformulation of the same theory. As a
byproduct of our analysis of the formalism we of\/fer an amendment of
the Dirac's bra-ket notation and we also show how its use clarif\/ies
the concept of covariance in time-dependent cases. Via an elementary
example we f\/inally explain why in certain quantum systems the
generator $H_{\rm (gen)}$ of the time-evolution of the wave functions
may {\em differ} from their Hamiltonian $H$.}


\Classification{81Q10; 47B50}

\renewcommand{\thefootnote}{\arabic{footnote}}
\setcounter{footnote}{0}

\section{Introduction}\label{secI}

In contrast to classical mechanics where various formulations of the
theory abound, there exist not too many alternative formulations of
{\em quantum} theory. Moreover, most of their lists (to be found
mostly in textbooks and just rarely in review papers like~\cite{Styer}) pay attention just to the history of the subject,
creating an impression that the formulation of quantum theory does
not lead to any interesting theoretical developments.

Ten years ago this false impression has been challenged by Bender
and Boettcher~\cite{BB} who surprised the physics community by a
numerically supported conjecture that quite a few one-dimensional
quantum potentials $V(x)$ may generate bound states $\psi_n(x)$ with
real energies $E_n$ even when the potentials themselves are not
real. The conjecture seemed to contradict the current experience
with quantum mechanics. Using the traditional textbook
single-Hilbert-space (1HS) formulation of the theory~\cite{Messiah}
we usually employ the fact that the separable Hilbert spaces are all
unitarily equivalent. This allows us to restrict attention, say, to
the exceptional and most user-friendly representations $\ell_2$ or
${\mathbb L}^2({\mathbb R})$ of the Hilbert space of states (cf. Appendix~\ref{appendixA} for a few additional comments). On this background one can expect
a complexif\/ication of the spectra whenever the underlying
Hamiltonian becomes manifestly non-Hermitian. It is just this naive
belief which has been shattered by the Bender's and Boettcher's
illustrative Hamiltonians which all possessed, as an additional
methodical benef\/it, the most elementary and common form $H=p^2+V(x)$
of the sum of the kinetic and potential ``energies''.

Obviously, any imaginary component in $V(x)$ makes the latter
Hamiltonians with real spectra safely non-Hermitian in
${\mathbb L}^2({\mathbb R})$. This is a paradox which evoked an intensive
interest in the families of apparently unphysical Hamiltonians~$H$
the Hermiticity of which appeared broken in a theoretically
inappropriate but mathematically exceptional and friendly specif\/ic
representation of an abstract Hilbert space which will be denoted
here as ${\cal H}^{\rm (auxiliary)}$,
 \begin{gather}
 H \neq H^\dagger\qquad  {\rm in }\quad  {\cal
 H}^{\rm (auxiliary)}.
 \label{neherme}
 \end{gather}
The progress in this direction of research has been communicated in
dedicated conferences (cf.~their webpage~\cite{webpage} and
proceedings~\cite{proc} or, better, the recent compact review paper~\cite{Carl}). It became clear that in spite of the undeniable appeal
of the models where $E_n=$ real while $V(x) =$ complex one must
treat their ``non-Hermiticity'' (\ref{neherme}) as an ill-conceived
concept. The operators $H$ with real spectra have been reinterpreted
as Hermitian after an appropriate {\it ad hoc} redef\/inition of the
Hilbert space ${\cal H}$ of states. For this purpose one must merely
replace the original, false space ${\cal H}^{\rm (auxiliary)}$ by
another, physical space ${\cal H}^{\rm (standard)}$ of states with the
standard quantum-mechanical meaning.

A few relevant aspects of such a theoretical and conceptual
innovation will be discussed in our present paper. Our introductory
Section~\ref{sII} and Appendix~\ref{appendixA} recollect the main features of the
currently most popular two-Hilbert-space (2HS) formulation of such a
version of quantum theory\footnote{With the roots dating back to
Scholtz et al.~\cite{Geyer} or even to Dieudonne et al.~\cite{Dieudonne}.} which is based on the use of the so called
quasi-Hermitian or, better, cryptohermitian\footnote{I.e.,
non-Hermitian in ${\cal H}^{\rm (auxiliary)}$ but Hermitian in ${\cal
H}^{\rm (standard)}$.} representation of observables. In Subsection~\ref{uusII}, in particular, we return brief\/ly to our recent paper~\cite{timedeprd} where the 2HS formalism has been shown suitable for
the description of such quasi-Hermitian quantum systems which
require the use of manifestly time-dependent operators of
observables.

In Section~\ref{sIII} our present main result is described showing
that a thorough simplif\/ication of the theory can be achieved when
one replaces its 2HS formulation by a more appropriate
three-Hilbert-space (3HS) reformulation. A few subtleties of the
resulting generalized formalism (as well as of our present amended
notation conventions) are illustrated via an elementary
time-independent two-dimensional matrix solvable example in
Section~\ref{sIV}.

Section~\ref{sV} returns again to the results of~\cite{timedeprd}. We stress there that one of the most remarkable
applications of the innovated 3HS formalism can be found in the
perceivably facilitated covariant construction of certain
sophisticated generators of time evolution. For illustration we
return to the matrix example of Section~\ref{sIV} and we
re-analyze its time-dependent generalized version in Section~\ref{sVI}. We f\/irmly believe that the  3HS description of this and
similar examples can perceivably simplify our understanding of
paradoxes which may emerge in quasi-Hermitian models.

The summary of our results is provided in Section~\ref{sVII}.
Several apparently anomalous properties of the Bender's and
Boettcher's potentials are discussed there as an inspiration of
revisiting a~few less rigorous formulations of the f\/irst principles
of quantum theory. No real necessity of the changes of these general
principles themselves is encountered. Still, our present
modif\/ication of their implementation and of the related conventions
in notation appears both desirable and benef\/icial.

\section[Quantum Mechanics in 2HS formulation - a brief
recollection]{Quantum Mechanics in 2HS formulation -- a brief
recollection}\label{sII}

The practical use of phenomenological Hamiltonians $H$ which are
apparently non-Hermitian (cf.~(\ref{neherme})) does certainly
enhance the f\/lexibility of the constructive model-building
activities in physics even behind the framework of quantum theory
(cf.~\cite{Uwe}). It also broadens the space for {\em feasible}
applications of non-local models in a way exemplif\/ied by the
above-mentioned {\em complex} Bender--Boettcher toy potentials~\cite{Carl}. In similar cases, an attachment of the {\em doublet} of
Hilbert spaces ${\cal H}^{\rm (auxiliary)}$ and ${\cal H}^{\rm (standard)}$
to a single quantum system may make good sense.

\subsection{The hermitization of cryptohermitian observables}\label{vvsII}

One of the f\/irst applications of the apparently non-Hermitian
Hamiltonians with real spectra appeared in nuclear physics~\cite{Geyer}. The {\em correct physical interpretation} of the model
in ${\cal H}^{\rm (standard)}$ has been separated there from the {\em
facilitated calculations} of the spectrum in ${\cal
H}^{\rm (auxiliary)}$. Further physical application of the same method
appeared in Mostafazadeh's study of the free relativistic
Klein--Gordon equation \cite{KGali} which is traditionally introduced
in the Feshbach's and Villars' \cite{FV} unphysical representation
space ${\cal H}^{\rm (auxiliary)}={\mathbb L}^2({\mathbb R})\bigoplus
{\mathbb L}^2({\mathbb R})$. A reconstruction of the inner product has
been of\/fered as a means of recovering the consistent picture of
physics. The same approach avoiding spurious states or negative
probabilities has been extended to the f\/irst-quantized models of
massive particles with spin one~\cite{jakub}.

On theoretical level the 2HS reformulation of quantum theory might
look almost trivial. Still, the application of the idea to the
Bender's and Boettcher's elementary examples and the resolution of
some of the related puzzles took time \cite{BBJ}. Fortunately, the
theory seems to be clarif\/ied at present. Its key mathematical
feature lies in the Hamiltonian-dependent replacement of the spaces,
 \begin{gather}
 {\cal H}^{\rm (auxiliary)} \to {\cal H}^{\rm (standard)} .
 \label{expo}
 \end{gather}
The detailed description of its mathematical subtleties can be found
explained in the available literature. The innovative 2HS approach
to the description of pure states in a quantum system characterized
by an apparently non-Hermitian Hamiltonian can even be presented
using the standard 1HS language (cf., e.g., \cite{Ali}). In
such an approach it is only necessary to introduce a rather
complicated notation in which {\em the same} state is characterized
by {\em two different} Greek letters (say, $\Phi$ and $\Psi$ as
recommended in \cite{Ali}).

Some details of this convention are recollected and summarized in
Appendix~\ref{appendixA.2}. Here, let us only emphasize that we must remember
that although equation~(\ref{expo}) does not involve a~change of the
underlying vector space ${\cal V}$ itself, it {\em does} modify
the inner product in this space. Thus, we must introduce two
graphically dif\/ferent Dirac's bra-vector symbols associated with
the individual Hilbert spaces ${\cal H}^{\rm (auxiliary/standard)}$.
Of course, this enables us to restore the necessary physical
Hermiticity of our Hamiltonian,
 \begin{gather*}
 H = H^\ddagger\qquad  {\rm in }\quad  {\cal
 H}^{\rm (standard)} .
 \end{gather*}
In the other words, even if we start from a non-Hermitian model
(\ref{neherme}), we may update the correct physical form of the
Hilbert space and re-establish, thereby, the validity of all of
the standard postulates of quantum theory.

\subsection{Cryptohermitian Hamiltonians  in 2HS picture} \label{uusII}

Although we do not intend to accept the above 1HS notation
conventions in our present paper, we would still like to keep our
present paper self-contained. For this reason we added further
comments on the 1HS notation and postponed them to Appendix~\ref{appendixA}. The
main reason is that we are persuaded that the consequent 2HS
notation which makes an explicit use of the two spaces appearing
in~(\ref{expo}) is perceivably simpler and less confusing.

We have to admit in advance that {\em neither} of the two spaces
${\cal H}^{\rm (auxiliary)}$ and ${\cal H}^{\rm (standard)}$  of\/fers in fact
a conceptually fully satisfactory frame for wave functions of a
given quantum system. Indeed, the former space remains manifestly
unphysical while the work in the latter one requires the
construction and use of a Hamiltonian-dependent metric operator
$\Theta \neq I$. Still, our recent application~\cite{timedeprd} of
the 2HS ideas to models with a nontrivial dependence on time
re-demonstrated the mathematical strength as well as physical
productivity of the 2HS approach (cf.~Table~\ref{tabtwo}).


\begin{table}[h]
\caption{Concise summary of the extended 2HS notation as employed in~\cite{timedeprd}.} \label{pexp4}
\begin{center}
\begin{tabular}{||c|c|c|c||}
\hline \hline
 \bsep{1ex}\tsep{1ex} {\rm Hilbert space} &{\rm ket state} &{\rm its dual}
  &  {\rm its Hamiltonian}
 \\
 \hline
  \hline
 \bsep{1ex}\tsep{1ex} ${\cal H}^{\rm (auxiliary)}_{\rm (unphysical)}$  $\equiv$ ${\cal H}^{(A)}$
  & $|\Phi\kt$
  & $\br \Phi|$
  & $ H\neq H^\dagger$
  \\
 \hline
 \bsep{1ex}\tsep{1ex}   ${\cal H}^{\rm (standard)}_{\rm (physical)} \equiv {\cal H}^{(P)}$
   & $|\Phi\kt$ & $\br \Psi|=\br
 \varphi|\Omega$ &  $ H=H^\ddagger$
 \\
  \hline
  \hline
  \bsep{1ex}\tsep{1ex}  ${\cal H}^{\rm (auxiliary)}_{\rm (physical)} \equiv {\cal H}^{(A)}$
    & $|\varphi \kt
    =\Omega|\Phi\kt$
  & $\br \varphi|=\br
 \Phi|\Omega^\dagger$ &  $ \mathfrak{h}=\Omega H \Omega^{-1}
 =\mathfrak{h}^\dagger$
  \\
  \hline
  \hline
\end{tabular}
 \label{tabtwo}
\end{center}
\end{table}

Some of the key ideas of~\cite{timedeprd} were inspired by
the transparency of the notation as suggested in~\cite{knots}. The core of their ef\/f\/iciency lies in the
simultaneous use of {\em two} dif\/ferent basis sets in the {\em
same} friendly Hilbert space ${\cal H}^{\rm (auxiliary)}$ (denoted as
${\cal H}^{(A)}$ in~\cite{timedeprd}). This ef\/fectively separated
the original computing-frame role of this space from its other
role of a benchmark physical space. A certain invertible
non-unitary transformation $\Omega: {\cal H}^{(A)} \to {\cal
H}^{(A)}$ has been invented as formally connecting these two roles
of space ${\cal H}^{\rm (auxiliary)}$. Section~III of paper
\cite{timedeprd} could be consulted for more details. The
correspondence between these two roles is ref\/lected also by the
f\/irst and last row in Table~\ref{tabtwo}.

The clarity of the message mediated by Table \ref{tabtwo} is
weakened by the fact that our notation has been taken from~\cite{Ali} in spite of its being not too suitable for the
given purpose. Indeed, the comparison of Table \ref{tabtwo} with
the 1HS Table \ref{tabone} of Appendix~\ref{appendixA.2} shows that the
separation of the two bases is not well ref\/lected by the notation.
The necessary use of the third reserved Greek letter $\varphi$
representing the same state only enhances the danger of confusion.
A more thoroughly amended version of the Dirac's notation is to be
of\/fered in the next section.

\section{Quantum Mechanics in 3HS formulation}\label{sIII}

During the proofreading of the text of~\cite{timedeprd} we
imagined that it of\/fers a slightly confusing picture of
cryptohermitian quantum systems, especially due to the use of the
imperfect 2HS notation as sampled here in Table~\ref{tabtwo}. As
we already noticed, it is rather unfortunate that this notation
employs {\em three} dif\/ferent Greek letters (viz., $\Phi$, $\Psi$
and $\varphi$) representing {\em the same} physical state. In
addition, this notation also introduces a strange asymmetry
between the two Hilbert spaces~${\cal H}^{(A)}$ and ${\cal
H}^{(P)}$.

There is in fact no reason why the former one should be treated as
a single Hilbert space because its underlying vector space ${\cal
V}$ is in fact being equipped with the two dif\/ferent inner
products. This is also the driving idea of our present proposal of
transition from 2HS to 3HS language. Its mathematical background
is virtually trivial as it makes merely use of the well known {\em
formal unitary equivalence} between {\em any two} (separable)
Hilbert spaces. Once applied to the two physical spaces of
Table~\ref{tabtwo}, we may declare the parallel mathematical and
physical equivalence between the second and the third item of this
Table, i.e., between the standard and auxiliary {\em physical}
Hilbert spaces {\em even if they cease to share the underlying
vector space} ${\cal V}$.

The main advantage of the resulting 3HS separation of the
constructive def\/initions of the latter two representations of the
Hilbert space lies in the possibility of the decoupling of the
underlying {\em vector} spaces,
 \begin{gather}
 {\cal V}={\cal V}_{{\cal H}^{\rm (standard)}_{\rm (physical)}} \neq
 {\cal V}_{{\cal H}^{\rm (new\ auxiliary)}_{\rm (physical)}}:={\cal W}
 \label{exnativ}
 \end{gather}
(cf.~Appendix~\ref{appendixA} for notation). In its turn, such a new,
3HS-specif\/ic freedom (\ref{exnativ}) enables us to get rid of the
extremely unpleasant nontriviality of the metric {\em also} in the
latter, {\em physical} Hilbert space,
 \[
 \Theta^{\rm (new\ auxiliary)}_{\rm (physical)}=I.
 \]
It is encouraging to see that the only price to be paid for this
3HS freedom lies just in the (non-unitary) generalization of the
mapping $\Omega$ which will now be acting between the {\em two
different} vector spaces,
 \[
 \Omega:\ \ {\cal V}\,\longrightarrow\,{\cal W} .
 \]
A more detailed analysis of some other consequences of the new
perspective may be found in Appendix~\ref{appendixB} below.

\subsection{3HS formulism}\label{uusIII}

In the  Bender--Boettcher-type bound-state  models where $H \neq
H^\dagger$ in ${\cal  H}^{\rm (auxiliary)}$ it proved convenient to
factorize their nontrivial, non-Dirac metric in ${\cal
H}^{\rm (standard)}$,
either in the form $\Theta = {\cal CP}$ (where ${\cal P}$ is parity
and ${\cal C}$ represents a charge \cite{Carl}) or in the form
$\Theta = {\cal PQ}$ (where ${\cal Q}$ is quasiparity~\cite{SIGMA}).
Unfortunately, after we turn attention to the other quantum systems
with the scattering-admitting  Hamiltonians  $H=T+V$ \cite{Cannata},
we discover that the construction of an appropriate metric $\Theta
\neq I$ only remains feasible for certain extremely elementary
models of dynamics~\cite{Jones}.

In this context, our present 3HS formulation of Quantum Mechanics
found one of its secondary sources of inspirations in the
possibility of a return from $\Theta^{\text{(non-Dirac)}}$ to
$\Theta^{\rm (Dirac)}$. This indicates that the second (in principle,
extremely complicated but still norm- and inner-product
preserving) update of the physical Hilbert space
 \begin{gather}
 {\cal H}^{\rm (standard)}_{\text{(non-Dirac)}}   \longrightarrow   {\cal H}^{\rm (new\ auxiliary)}_{\rm (Dirac)}
  :=
 {\cal H}^{(T)}
 \label{exponate}
 \end{gather}
proves desirable and very natural. It can also be read as an
introduction of the {\em third} Hilbert space ${\cal H}^{(T)}$.
Thus, equation~(\ref{exponate}) complements equation~(\ref{expo}) above.

All the necessary details and formulae can be found again shifted
to Appendix~\ref{appendixB}. Here, let us only summarize that the symbol ${\cal
H}^{(T)}$ representing the third space in equation~(\ref{exponate})
will be accompanied, in what follows, by the other two
abbreviations representing the {\em first} and the {\em second}
Hilbert spaces
 \[
 {\cal H}^{\rm (auxiliary)}_{\rm (unphysical)}:={\cal H}^{(F)} ,
 \qquad
 {\cal H}^{\rm (standard)}_{\rm (physical)}:={\cal H}^{(S)}
  \]
of Table~\ref{tabtwo}, respectively. In summary we can now
recommend that the dif\/ferences in mappings between our three
dif\/ferent Hilbert spaces ${\cal H}^{(F,S,T)}$ can be very easily
ref\/lected by the dif\/ferences in a rationalized Dirac notation
where just the graphical form of the bras and kets will be varied.

This will enable us to correlate the graphical form of the bras
and kets with the three individual Hilbert spaces. At the same
time, the same letter (say, $\psi$) will {\em always} represent
{\em the same} physical state. Preliminarily, this pattern of
notation is summarized in Table~\ref{tabtri}. Multiple parallels
with Table~\ref{tabtwo} can be noticed here.

\begin{table}[h]
\caption{3HS notation: A given state $\psi$ in three alternative
representations.} \label{pxp4}
\begin{center}
\begin{tabular}{||c|c|c|c||}
\hline \hline
\bsep{1ex}\tsep{1ex} {\rm Hilbert space} &{\rm ket-vector}&{\rm bra-vector}
 & {\rm  norm squared}
 \\
 \hline
  \hline
\bsep{1ex}\tsep{1ex}
 ${\cal H}^{(F)}_{\rm (friendly)}$& $|\psi\kt \in {\cal V}$ & $\br \psi|$ &
 $ \br \psi|\psi \kt$
 \\
 \hline
\bsep{1ex}\tsep{1ex}
 ${\cal H}^{(S)}_{\rm (standard)}$ & $|\psi\kt \in {\cal V}$ & $\br \br \psi|= \pbr
 \psi|\Omega$ & $\br \br \psi|\psi \kt
  =\br \psi|\Omega^\dagger\Omega|\psi \kt$
 \\
  \hline
\bsep{1ex}\tsep{1ex}
 ${\cal H}^{(T)}_{\rm (textbook)}$& $|\psi \pkt  =\Omega|\psi\kt \in {\cal W}$ & $\pbr \psi|=\br
 \psi|\Omega^\dagger$ & $ \pbr \psi|\psi \pkt =\br \psi|\Omega^\dagger\Omega|\psi \kt$
 \\
 \hline
 \hline
\end{tabular}
 \label{tabtri}
\end{center}
\end{table}

\subsection[Metric-eliminating transformation  $\Omega$]{Metric-eliminating transformation  $\boldsymbol{\Omega}$}

The explicit use of mappings between Hilbert spaces is quite
common in textbooks~\cite{Messiah} where a~{\em unitary} map
(e.g., Fourier transformation $\Omega$) produces the
correspondence. In the 3HS context the same transition is being
postulated,
 \begin{gather}
 |\psi\kt \in {\cal H}^{(F,S)}\
 \Longrightarrow \
 |\psi\pkt \equiv \Omega\, |\psi\kt \in {\cal H}^{(T)} .
 \label{ket}
 \end{gather}
Nevertheless, the majority of the nontrivial aspects of the
present three-Hilbert-space approach to quantum models will only
emerge when $\Omega$ ceases to be norm-preserving (let us still
say unitary). In such a setting the two physical spaces of states
${\cal H}^{(S)}$ and ${\cal H}^{(T)}$ are accompanied by their
unphysical partner ${\cal H}^{(F)}$. This partnership can already
be perceived as aiming at a~reformulation of quantum theory.
Another hint lies in the isospectrality of the Hamiltonians, of
which $
 \mathfrak{h}$ acts in ${\cal H}^{(T)}$ while
 $H \equiv \Omega^{-1}\mathfrak{h} \Omega$ acts in
  ${\cal H}^{(F)}$ or in $ {\cal H}^{(S)}$.
This opens a constructive possibility of the choice of a
Hamiltonian which is {\em allowed} to be non-Hermitian (cf.\
equation~(\ref{neherme})).

{\it In nuce}, our present main technical trick is that in place
of the {\em unitary} transformation of spaces (\ref{exponate})
(i.e., the second option in equation~(\ref{ket})) we intend to achieve
the same goal indirectly, by means of the technically less
dif\/f\/icult and {\em non-unitary} transition between the other two
Hilbert spaces [i.e., via the f\/irst option in equation~(\ref{ket})],
 \begin{gather*}
 {\cal H}^{\rm (auxiliary)}_{\rm (Dirac)}   \big[={\cal H}^{(F)}\big ]
 \to {\cal H}^{\rm (standard)}_{\rm (Dirac)}   \big[={\cal H}^{(T)}\big
 ]
 \end{gather*}
both of which are equipped with the same and, namely, trivial
Dirac metric. The key features of the latter idea may be read out
of the parallelled Tables~\ref{tabtwo} and~\ref{tabtri}.

A  detailed inspection of Table \ref{tabtri} reveals  the
coincidence of the kets in the ``$F$'' and ``$S$'' doublet. The mapping
between the respective ``$S$'' and ``$T$'' Hilbert spaces ${\cal H}^{(S)}$
and ${\cal H}^{(T)}$ preserves the inner product and is, in this
sense, unitary,
 \[
 \pbr\psi'|\psi\pkt\ =\bbr\psi'|\psi\kt
   .
\]
Equivalent physical predictions will be obtained in {\em both} of
the latter spaces. The third pair of the ``Dirac-metric'' spaces with
$\Theta^{\rm (Dirac)}=I$ and  superscripts ``$F$'' and ``$T$'' shares the form
of the Hermitian conjugation.

In such a balanced scheme the space ${\cal H}^{(T)}$ is slightly
exceptional. Not only by its full compatibility with the standard
textbooks on quantum physics but also by its role of an extremely
computing-unfriendly (i.e., practically inaccessible) representation
space. In both of these roles its properties are well exemplif\/ied by
the overcomplicated fermionic Fock space which occurred in the
above-mentioned nuclear-physics context~\cite{Geyer}. Summarizing,
all of the three spaces in Table~\ref{tabtri} can be arranged, as
vector spaces, in the following triangular ket-vector pattern
$$
  \begin{array}{c}
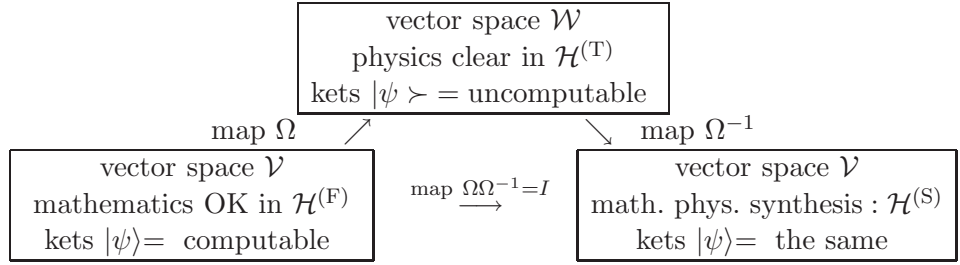
    \begin{array}{|c|}
 \hline
  \ {\rm vector\ space} \
 {\cal W}\ \\
  \ {\rm physics\ clear\ in} \
 {\cal H}^{\rm (T)}\ \\
    {\rm kets}\ |\psi\pkt
  {\rm  = uncomputable} \ \\
 \hline
 \end{array}
\\ {\rm map}\
 \Omega \ \ \ \   \nearrow \ \  \  \ \ \ \ \ \
 \ \ \ \ \ \ \ \
\ \ \ \ \   \searrow  \ \ {\rm map}\  \Omega^{-1}\\
 \begin{array}{|c|}
 \hline
  \ {\rm vector\ space} \
 {\cal V}\ \\
 \ {\rm mathematics\ OK\ in\ }
 {\cal H}^{\rm (F)}\ \\
    {\rm kets}\ |\psi\kt
  {\rm =\ computable}\  \\
  \hline
 \end{array}\ \ \
 \stackrel{{\rm map}\ \Omega  \Omega^{-1}=I}{ \longrightarrow }
 \ \
 \begin{array}{|c|}
 \hline
  \ {\rm vector\ space} \
 {\cal V}\ \\
 {\rm math.\ phys.\ synthesis}:
 {\cal H}^{\rm (S)} \\
  {\rm kets}\ |\psi\kt
  {\rm =\ the \ same} \ \\
 \hline
 \end{array}
\\
\end{array}
$$
In parallel we have to study the bras. After the transition to the
conjugate vector spaces of functionals the above-indicated pattern
gets modif\/ied as follows,
$$
  \begin{array}{c}
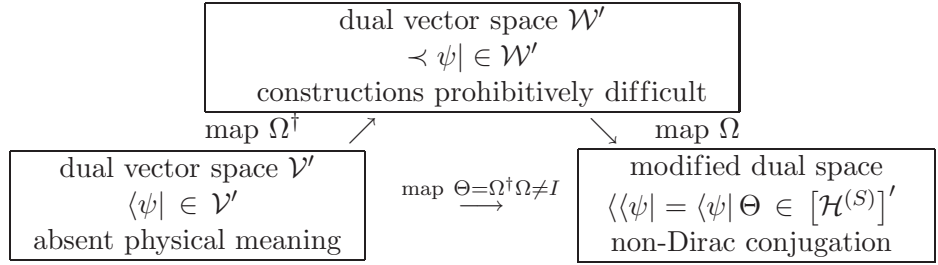
    \begin{array}{|c|}
 \hline
  \ {\rm dual\ vector\ space} \
 {\cal W}'\ \\
 %
  \pbr \psi| \in
 {\cal W}'
 \\
  \ \ \ \ {\rm constructions \ prohibitively\ dif\/f\/icult\ } \ \\
 \hline
 \end{array}
\\  {\rm map}\
 \Omega^\dagger \ \ \ \
 \nearrow \ \  \  \ \ \ \ \ \
 \ \ \ \ \ \ \ \
\ \ \ \ \  \searrow \ \ \  {\rm map}\ \Omega \\
 \begin{array}{|c|}
 \hline
  \ {\rm dual\ vector\ space} \
 {\cal V}'\ \\
 %
  \br \psi|\,\in\,
 {\cal V}'\ \\
 \
  {\rm  absent\ physical\ meaning}\    \\%
  \hline
 \end{array}\ \ \
 \stackrel{ {\rm map}\ \Theta=\Omega^\dagger\Omega\neq I }{ \longrightarrow }
 \
 \begin{array}{|c|}
 \hline
  \ {\rm modif\/ied\ dual\  space} \
  \\
 %
 \bbr \psi|=\br \psi|\,\Theta\,\in\,
 \left [{\cal H}^{(S)}\right ]'\
 \\ \ \
  \text{non-Dirac\ conjugation}\ \ \  \\%
 \hline
 \end{array}
\\
\end{array}
$$

\section{An elementary illustrative example}\label{sIV}

In the second row of Table \ref{tabtwo} one f\/inds the condition of
the Hermiticity of our Hamiltonian $\mathfrak{h}$ in~${\cal
H}^{(T)}$. The pullback of this condition to ${\cal H}^{(F)}$ gives
 \begin{gather}
 H^\dagger = \Theta H \Theta^{-1} , \qquad
 \Theta=\Omega^\dagger\Omega=\Theta^\dagger > 0
 \label{quha}
 \end{gather}
so that our upper-case Hamiltonian is similar to its Hermitian
conjugate or, in the terminology of~\cite{Geyer,Dieudonne}, it
is quasi-Hermitian.

One of the most elementary illustrative examples of a
quasi-Hermitian $H$ has been proposed by Mostafazadeh
\cite{timedep}. In a toy space  ${\cal H}^{(F)}$ which is just
two-dimensional, this Hamiltonian is represented by the
two-dimensional matrix
 \begin{gather}
 H=H^{(AM)}(r,\beta)=\left (
 \begin{array}{cc}
 0&r\,e^{{\rm i}\beta}\\
 r^{-1}\,e^{-{\rm i}\beta}&0
 \end{array}
 \right )
 \label{oho}
 \end{gather}
which is strictly two-parametric, $\beta\in (0,2\pi)$ and $r \in
{\mathbb R} \setminus \{0\}$. Its condition of quasi-Hermiti\-ci\-ty~(\ref{quha})
can be read as four linear homogeneous algebraic equations
determining the matrix elements of all the eligible positive
def\/inite matrices $\Theta=\Theta^{(AM)}$. The general solution of
these equations
 \begin{gather}
 \Theta^{(AM)}=f \cdot \Theta_Z ,
 \qquad \Theta_Z=
 \left (
 \begin{array}{cc}
 1&r\,e^{{\rm i}\beta}\cos Z \\
 r e^{-{\rm i}\beta}\cos Z&r^2
 \end{array}
 \right )
 \label{ures}
 \end{gather}
depends on two new parameters or, if we ignore the overall factor
$f$, on $Z \in (0,2\pi)$.

Any other observable quantity must be represented by the operator
$\Lambda$ which is also quasi-Hermitian with respect to the same
metric,
 \begin{gather}
 \Theta \Lambda  =
 \Lambda^\dagger \Theta
 .
 \label{berese}
 \end{gather}
The inverse problem of specif\/ication of all of the eligible
$\Lambda$s is, in our schematic example, easily solvable,
 \begin{gather}
 \Lambda=
 \left (
 \begin{array}{cc}
 a&p e^{{\rm i}\beta} \\
 q e^{-{\rm i}\beta}&d
 \end{array}
 \right ) .
 \label{family}
 \end{gather}
In this family of solutions the range of the four real parameters
$a$, $p$, $q$ and $d$ is only restricted by the inequality
$(a-d)^2>4pq$ which guarantees the reality of both the observable
eigenvalues and by the single nontrivial constraint resulting from
quasi-Hermiticity equation,
 \begin{gather}
 p=qr^2+(a-d)r\cos Z .
 \label{connie}
 \end{gather}
Once we f\/ix $Z$ and $f>0$ in equation~(\ref{ures}), an illustrative
factorization $\Theta_Z=\Omega^\dagger_Z \Omega_Z$ of the metric
can be performed, say, in terms of triangular matrices,
 \[
 \Omega_Z=
 \left (
 \begin{array}{cc}
 1&r e^{{\rm i}\beta}\cos Z \\
 0&r \sin Z
 \end{array}
 \right ) ,
 \qquad
 \Omega^{-1}_Z=
 \left (
 \begin{array}{cc}
 1&-e^{{\rm i}\beta}\cot Z \\
 0&1/(r \sin Z)
 \end{array}
 \right )
 .
 \]
This def\/inition specif\/ies, f\/inally, the family of eligible
selfadjoint $Z$-dependent Hamiltonians
 \[
 \mathfrak{h}^{(AM)} \sim \mathfrak{h}_Z=
 \left (
 \begin{array}{cc}
 \cos Z&e^{{\rm i}\beta}\sin Z \\
 e^{-{\rm i}\beta}\sin Z &-\cos Z
 \end{array}
 \right )
  \]
def\/ined by pullback to ${\cal H}^{(T)}$ and isospectral with
$H^{(AM)}$. The same pullback mapping must be also applied to the
second observable $\Lambda$ of course.

Marginally, let us mention that in many items of the current
literature (well exemplif\/ied by~\cite{KGali}) the factorization of
$\Theta$ is only being made in terms of very special Hermitian and
positive def\/inite mapping operators $\Omega_{\rm (herm)} =
\Theta^{1/2}$. In this context our illustrative example shows that
the choice of the special $\Omega_{\rm (herm)}$ is just an arbitrary
decision rather than a necessity dictated by the mathematical
framework of quantum theory. We shall see below that a consistent
treatment of the ambiguity of our choice of $\Omega$ is in fact very
similar to the treatment of the ambiguity encountered \cite{Geyer}
during the assignment of the metric $\Theta$ to a given Hamiltonian~$H$.

\section{Covariant construction of the generator of time-evolution}\label{sV}

In the  3HS formulation of quantum physics we may start building
phenomenological models inside any item of the triplet ${\cal
H}^{(F,S,T)}$. Still, in a way noticed in~\cite{timedep} and
worked out in~\cite{timedeprd} this freedom may be lost when one
decides to admit also the models where the parameters which def\/ine
the system and its properties (i.e., say, the parameters in the
quasi-Hermitian Hamiltonian~$H$ of equation~(\ref{oho}) and/or in
another observable $\Lambda$ given by equation~(\ref{family})) become
allowed to vary with time. Typically, this time-dependence may
involve not only the external forces which control the system but
also, in principle, the related measuring equipment.

In such an overall setting one has to imagine that the constraints
imposed upon the assignment of a suitable metric $\Theta$, say, to a
given $H$ {\em and} $\ \Lambda$ may prove impossible. In the
language of our illustrative example this danger has been
illustrated in \cite{timedep} where in  illustrative example
(\ref{oho}) the so called quasistationarity condition $\Theta \neq
\Theta(t)$  has been shown to imply that $r \neq r(t)$ and $\beta
\neq \beta(t)$. The mathematical reason was easy to f\/ind since the
quasi-stationary time-dependent generalization of the set of
constraints (\ref{quha}) and (\ref{berese}) proved overcomplete.

\subsection{Time-dependence and Schr\"{o}dinger equations}\label{usVI}

In \cite{timedeprd} we have shown the reasons why the same
overrestrictive role is played by the overcompleteness also in the
generic, model-independent quasi-stationary scenario. In the
opposite direction, our study \cite{timedeprd} recommended to relax
the quasi-stationarity constraint as too artif\/icial. Then, the 3HS
formalism proved applicable in its full strength. Now, we intend to
show that it restricts the form of the time-dependence of the
operators of observables as weakly as possible.

We shall again ``teach by example'' and demonstrate our statement via
the same schematic example as above. Still, we have to recollect the
basic theory f\/irst. Thus, we start from the exceptional space ${\cal
H}^{(T)}$ which of\/fers the absence of all doubts in the physical
interpretation of any 3HS model. In particular, the time-evolution
will be controlled by the textbook Schr\"{o}dinger equation in
${\cal H}^{(T)}$,
\[
 {\rm i}\p_t|\psi(t)\pkt = {\mathfrak{h}}\,|\psi(t)\pkt  .
 \label{SEti}
\]
This is fully compatible with the textbook concepts of the
measurement~\cite{Messiah}  based on the idea that the complete
physical information about  a given system prepared in the so called
pure state is compressed in its time-dependent wave function
$|\psi(t)\pkt$.

Secondly, in ${\cal H}^{(T)}$ there are also no doubts about the
standard postulate that all of the other measurable characteristics
of the system in question are obtainable as eigenvalues or mean
values of the other operators of observables exemplif\/ied here, for
the sake of simplicity, by $\lambda=\Omega \Lambda \Omega^{-1}$.
The analysis of this aspect of the problem is postponed to the next
section here.

\subsection{Models with time-dependent Hamiltonians}\label{wwsVI}

In a preparatory step of the study of the time-evolution problem let
us just be interested in the single observable (viz., Hamiltonian)
and let us admit that a manifest time-dependence occurs in all of
the operators,
 \[
 \mathfrak{h}(t)=\Omega(t) H(t) \Omega^{-1}(t) .
 \label{mapped}
 \]
An easier part of our task is to write the time-dependent
Schr\"{o}dinger equation  in ${\cal H}^{(T)}$,
 \[
 {\rm i} \partial_t |\varphi(t) \pkt   =  \mathfrak{h}(t)
 |\varphi(t) \pkt .
 \label{timeq}
 \]
Its formal solution  $ |\varphi(t) \pkt   =
u(t) |\varphi(0) \pkt$ employs just the usual evolution operator,
 \begin{gather}
 {\rm i}\partial_t u(t)=\mathfrak{h}(t) u(t)
 \label{seh}
 \end{gather}
which is unitary in ${\cal H}^{(T)}$ so that we can conclude that
the norm of the state in question remains constant,
 \[
 \pbr \varphi(t)  |
 \varphi(t)\pkt=
 \pbr \varphi(0)  |
 \varphi(0)\pkt .
 \]
In the next step we recollect all our previous considerations and
def\/ine $|\varphi(t)\kt=\Omega^{-1}(t)  |\varphi(t) \pkt$ and
$\br \br \varphi(t) |=\pbr \varphi(t) | \Omega(t)$. We are then
able to distinguish between the two formal evolution rules in the
physical space ${\cal H}^{(S)}$. One of them controls the evolution
of kets,
 \[
 |\varphi(t)\kt=U_R(t)  |\varphi(0)\kt ,\qquad
 U_R(t)=\Omega^{-1}(t) u(t) \Omega(0)
 \]
while the other one (written here in its ${\cal H}^{(F)}$-space
conjugate form) applies to ketkets $|\cdot \kkt  \equiv ( \bbr
\cdot|)^\dagger$,
 \[
 |\varphi(t)\kt \kt=U_L^\dagger(t)  |\varphi(0)\kt \kt ,\qquad
 U_L^\dagger(t)=\Omega^\dagger(t) u(t)
 \left [\Omega^{-1}(0)\right ]^\dagger .
 \]
The pertaining dif\/ferential operator equations for the two (viz.,
right and left) evolution operators read
 \[
 {\rm i}\partial_t U_R(t)=
 -\Omega^{-1}(t)
 \left [{\rm i}\partial_t\Omega(t)
 \right ]  U_R(t)+H(t)  U_R(t)
 \]
and
 \[
 {\rm i}\partial_t U_L^\dagger(t)=
 H^\dagger(t)\, U_L^\dagger(t)
 +
 \big [{\rm i}\partial_t\Omega^\dagger(t)
 \big ]
 \big[
 \Omega^{-1}(t)
 \big]^\dagger  U_L^\dagger(t),
 \]
respectively. They are obtained, very easily, by the elementary
insertions of the def\/initions. Inside ${\cal H}^{(S)}$, the
conservation of the norm $ \br \br \varphi(t) | \varphi(t)\kt $
of states is re-established and paralleled by the same phenomenon
inside  ${\cal H}^{(T)}$. One has to solve, therefore, the two
equations which form the doublet of non-Hermitian partners of the
standard single evolution equation~(\ref{seh}) in the third and most
computation-friendly space ${\cal H}^{(F)}$.

We may conclude that the conservation of the norm of the states
which evolve with time in~${\cal H}^{(S)}$ becomes a trivial
consequence of the unitary equivalence of the model to its image in
${\cal H}^{(T)}$ (cf.~the explicit formulae in Table~\ref{tabtwo}). One can
also recommend the abbreviation $\dot{\Omega}(t)\equiv
\partial_t\Omega(t)$ which enables us to introduce the time-evolution
generator in ${\cal H}^{(F)}$,
 \[
 H_{\rm (gen)}(t)=H(t) -{\rm i}\Omega^{-1}(t)
 \dot{\Omega}(t) .
 \]
Its most remarkable feature is that it remains the same for {\em
both} the time-dependent Schr\"{o}dinger equations in ${\cal
H}^{(S)}$,
 \begin{gather}
 {\rm i}\partial_t|\Phi(t)\kt
 =H_{\rm (gen)}(t) |\Phi(t)\kt ,
 \label{SEA}\\
 {\rm i}\partial_t|\Phi(t)\kt \kt
 =H_{\rm (gen)}(t) |\Phi(t)\kt \kt .
 \label{SEbe}
 \end{gather}
A virtually  equally remarkable feature of the operator $
H_{\rm (gen)}(t)$ is that it {\em ceases} to be an ele\-men\-tary
observable in ${\cal H}^{(S)}$~\cite{z}. This is very natural of
course. The reason is that by our assumption the time-dependence
of the system ceases to be generated solely by the Hamiltonian
(i.e., energy-operator). Indeed, the manifest time-dependence of
the other operators of observables represents an independent and
equally relevant piece of input information about the dynamics.

\section{The two-by-two example revisited}\label{sVI}

The core of our present message is that after the change of the
representation of the Hilbert space of states  ${\cal H}^{(T)}\to
{\cal H}^{(S)}$  one should not insist on the survival of the
time-dependent Schr\"{o}dinger equation in its usual form where the
Hamiltonian acts as the generator of time shift. We have shown that
the doublet of equations~(\ref{SEA}) + (\ref{SEbe}) must be used instead.
Such a replacement opens the space for the consistent use of a broad
class of  metrics (i.e., spaces ${\cal H}^{(S)}$) which vary with
time!

\looseness=1
The consequences of the new freedom in $\Theta=\Theta(t)$ may be
illustrated again on our elementary two-by-two model (\ref{oho})
where both the parameters $ r=r(t) \in {\mathbb R}\setminus \{0\}$ and
$\beta=\beta(t) \in {\mathbb R}$ may now {\em arbitrarily} depend on
time. In such an exemplif\/ication the explicit time-dependence of the
metric
 \[
 \Theta(t)=f(t) \cdot \Theta_{Z(t)} ,
 \qquad \Theta_{Z(t)}=
 \left (
 \begin{array}{cc}
 1&r(t) e^{{\rm i}\beta(t)}\cos {Z(t)}\\
 r(t) e^{-{\rm i}\beta(t)}\cos {Z(t)}&r^2(t)
 \end{array}
 \right )
 \]
appears to have {\em two} independent sources. Firstly it results
from the direct transfer of the mani\-fest time-dependence from the
Hamiltonian $H(t)$. Secondly, a new source of time-dependence enters
via the new free function $Z(t)$ of time. Obviously, its
time-dependence may be read {\em either} as responsible for the
time-dependence {\em freedom} in the metric $\Theta(t)$ {\em or} as
a {\em consequence} of the time-dependence of the second observable
$\Lambda(t)$ which is, in principle, independently {\em prescribed}.

Thus, via the existence and role of the function $Z(t)$ of time our
example illustrates that the time-variation of {\em both} $H(t)$ and
$\Lambda(t)$ must be read as the two components of an input, {\it
external} information about the dynamics of our system. This
information cannot be contradicted by any additional constraints
upon the metric and, in this sense, this information restricts the
freedom in our consistent choice of the metric $\Theta(t)$.

For the f\/irst time the latter connection between metric $\Theta(t)$
and observables $H(t)$ and $\Lambda(t)$ has been noticed by the
authors of~\cite{Fring} and formulated in~\cite{timedep}.
One can conclude that in similar situations we are not allowed to
impose additional conditions upon $\Theta(t)$ so that the inner
products in our Hilbert space ${\cal H}^{(S)}$ vary with time in
general.

\looseness=1
This looks like a paradox but its essence is suf\/f\/iciently clearly
illustrated by our example where, due to its simplicity, all the
one-to-one correspondence between the variations of observables and
the metric is mediated by the mere single function $Z(t)$. The
time-dependence of $Z(t)$ enters the game via an external
specif\/ication of some particular form of the observable~$\Lambda(t)$. Nevertheless, we already know that this observable
must have the above-mentioned four-parametric structure
(\ref{family}). Its four real parameters $a(t)$, $p(t)$, $q(t)$ and
$d(t)$ {\em must be} mutually coupled by the time-dependent version
of constraint (\ref{connie}),
 \[
 p(t)=q(t)r^2(t)+[a(t)-d(t)]r(t)\cos Z(t) .
 \]
This formula may easily be reinterpreted as an implicit {\em
def\/inition} of the originally arbitrary function,
 \[
 Z(t)={\rm arccos}\,\frac{p(t)-q(t)r^2(t)}{[a(t)-ad(t)]r(t)} .
  \]
In our illustrative two-by-two example we may conclude that the {\em
input} information about the time-dependence of $\Lambda(t)$
represents and independent and relevant contribution to the
time-dependence of the metric. Thus, even if we just parallel the
construction of Section~\ref{sIV} and employ the triangular
factorization of the metric with,
\[
 \Omega_{Z(t)}=
 \left (
 \begin{array}{cc}
 1&r(t) e^{{\rm i}\beta(t)}\cos Z(t) \\
 0&r(t) \sin Z(t)
 \end{array}
 \right )
\]
the dif\/ference between the two operators $H(t)$ and $H_{\rm (gen)}(t)$
(representable by a slightly cumbersome closed formula) will not
vanish in general.

\section{Summary}\label{sVII}

The main message of this text is a conf\/irmation of the full
compatibility of several recent applications of quantum theory
with its standard form known to us from textbooks. This being
said, the present innovative 3HS formulation of the theory can be
classif\/ied as lying very close, in its spirit if not dictum, to
the one of\/fered by Scholtz et al.~\cite{Geyer}. Here we just showed
that there is no need to feel afraid of the use  (i) of the
3HS-extended Dirac notation and (ii) of the concept of covariance
in the applications of the theory to the systems where observables
happen to be manifestly time-dependent.

In the literature, depending on the respective authors, the
applications in question carry the respective names of
quasi-Hermitian quantum mecha\-nics (cf.\ the oldest two proposals
of\/fered, practically independently, in mathematics
\cite{Dieudonne} and physics~\cite{Geyer}), ${\cal PT}$-symmetric
quantum mecha\-nics (= the best advertised Carl Bender's trademark~\cite{Carl}), pseudo-Hermitian quantum mechanics (= perceivably
more general concept dating back to M.G.~Krein and given second
life by Ali Mostafazadeh~\cite{Ali}, with possible applications
reaching far beyond quantum mechanics) or cryptohermitian quantum
mechanics (by my opinion, the most explanatory name proposed, only
very recently, by Andrei Smilga).

Although {the formalism} of quantum theory is most often formulated
in the language of functional analysis and linear algebra (LA, using
Hilbert spaces, etc), the concrete {\em realizations} of the
operators of observables (i.e., Hamiltonians, etc) are very often
chosen as fairly elementary dif\/ferential operators. The f\/irst two
physicists who noticed a deep mathematical relationship between
these two ingredients in phenomenology were probably Bender and Wu~\cite{BW}. Almost forty years ago they discovered that for the
quartic anharmonic oscillator of coupling $g>0$ {\em all} the
spectrum of bound state energies $E_0(g) < E_1(g) < \cdots$ is given
by the {\em single analytic function} of {\em complex} $g$.

\looseness=1
Although this amazingly close correspondence between the analytic
and LA aspects of bound states still eludes a full appreciation, its
other manifestation has been revealed by Bender and Boettcher ten
years ago~\cite{BB}. They conjectured that for a family of
elementary and analytic {\em complex} oscillator potentials {\em
all} the spectrum of bound state energies $E_0 < E_1 < \cdots$ is
{\em strictly real}. Three years later, the {purely analytic
aspects} have been shown decisive in a rigorous proof of the reality
of the spectrum~\cite{DDT}. One year later, several consequences
have been drawn also in a 2HS reformulation of the underlying LA
formalism of quantum theory~\cite{BBJ}. An {\it ad hoc} operator of
metric $\Theta={\cal CP}$ has been introduced there. In LA language,
as a indirect consequence of the analyticity of the Bender's and
Boettcher's potentials, two non-equivalent Hilbert spaces, i.e., in
our present notation, ${\cal H}^{(F,S)}$ proved needed.

\looseness=-1
A counterintuitive limitation of the 2HS formalism to the models
with time-independent metrics $\Theta \neq \Theta(t)$ has been
revealed by A. Mostafazadeh \cite{timedep}. In \cite{timedeprd} we
removed this limitation via an introduction of the third Hilbert
space (in our present notation, of ${\cal H}^{(T)}$). Unfortunately,
we preserved the Mostafazadeh's 2HS notation which made the
resulting gem of the 3HS formalism clumsy.

In our present paper we introduced, therefore, an adequate
generalization of the traditional Dirac notation and  described the
updated 3HS formulation of quantum theory in full detail. We also
illustrated this formalism via an elementary solvable
two-dimensional matrix example. This example demonstrates some
details of the theory and of its applications. In particular, the
solvability of this  example underlines an easy nature of the
transition to the models with time-dependent metric.

\looseness=1
In the summary we would like to emphasize that the use of the
triplet of Hilbert spaces ${\cal H}^{(F,S,T)}$ and of the related
adapted Dirac notation can f\/ind its applicability in many
apparently dif\/ferent contexts ranging from computational and
variational nuclear physics \cite{Geyer} and analytic perturbation
theory \cite{Caliceti} through phenomenological f\/ield theory
\cite{Carl} up to the f\/irst quantization of  Klein--Gordon f\/ield
\cite{KGali} or Proca's f\/ield \cite{jakub}. For the nearest future
one could therefore encourage the use of the time-dependent metric
$\Theta(t)$ in all of these physically interesting contexts.

\appendix

\section{The standard Dirac notation and its shortcomings}\label{appendixA}

In some elementary introductions to Quantum Mechanics the abstract
concept of  Hilbert space~${\cal H}$ is being replaced by its
concrete representations, say, $\ell_2$ where the elements of the
space are identif\/ied with the ordered sets of complex numbers
arranged as column vectors. To each of these elements one then
associates a dual row vector obtained via the so called
Hermitian-conjugation operation (i.e., transposition plus complex
conjugation). Similarly, the analysis of some concrete physical
quantum systems can be based on the restriction of our attention to
the other concrete representations of ${\cal H}$ like, e.g., the
functional space ${\mathbb L}^2({\mathbb R})$ as mentioned in Section~\ref{secI} or the Klein--Gordon-equation space ${\cal
H}^{\rm (auxiliary)}={\mathbb L}^2({\mathbb R})\bigoplus {\mathbb L}^2({\mathbb R})$
as cited at the beginning of Section~\ref{sII}.

One of the most serious weaknesses of such a pedagogically
simplif\/ied approach is that whenever we start analyzing the
quasi-Hermitian models with property~(\ref{neherme}) we are forced
to turn attention to a modif\/ied inner product~\cite{Geyer}. This
means that in order to avoid confusion one has to replace the
concrete models like $\ell_2$ by the more general notion of the
abstract Hilbert space as given, say, by von Neumann.

\subsection[Difficulties with  variable  inner
products]{Dif\/f\/iculties with  variable  inner
products}\label{appendixA.1}

The term ``Hilbert space'' and its symbol ${\cal H}$ can be assigned
the precise mathematical meaning in several ways. One of the most
common def\/initions specif\/ies the abstract Hilbert space ${\cal H}$
as a vector space ${\cal V}={\cal V}_{\cal H}$ equipped with a
suitable inner product. By this ``product'' a complex number $c \in
{\mathbb C}$ is assigned to any pair of elements $a$ and $b$ of ${\cal
V}$. In the Dirac-inspired notation many authors write simply $c=\br
a|b\kt$ (cf., e.g., \cite{Ali}).

The main advantage of such a von Neumann-inspired approach is that
it does not require the knowledge of the more general concept of
Banach spaces and that it still enables us to overcome certain
mathematical dif\/f\/iculties in rigorous manner. The most serious ones
emerge when the vector space ${\cal V}$ ceases to be
f\/inite-dimensional. Then, the abstract inner-product spaces of this
type are required complete as metric spaces. A dual partner ${\cal
V}'$ of the vector space ${\cal V}$ itself is also very easily
def\/ined, in this language, as a set of bounded linear maps $\mu :
{\cal V} \to {\mathbb C}$. Furthermore, the requirement of the
self-duality of Hilbert space ${\cal H}$ can be reformulated as the
existence of a (by far not unique) isomorphism ${\cal T}$ between
vector spaces  ${\cal V}$ and ${\cal V}'$. The ``canonical''
isomorphism ${\cal T}^{\rm (Dirac)}: a \to \mu_a^{\rm (Dirac)}$ with $a \in
{\cal V}$ is usually introduced by the formula
$\mu_a^{\rm (Dirac)}(\cdot) =\br a | \cdot \kt$, i.e., it clearly and
explicitly depends on our choice of the initial inner product~$\br
\cdot | \cdot \kt$.

Whenever we decide to work with the single and f\/ixed inner product
$\br \cdot | \cdot \kt$, we are allowed to identify the elements $a$
of ${\cal V}$ with the Dirac's ket-vectors $| a \kt$. The parallel
identif\/ication of the linear functionals $\mu_b^{\rm (Dirac)}(\cdot)\in
{\cal V}'$ with the Dirac's bra-vectors $\br b |$ remains equally
straightforward but it will not survive a change of the inner
product. This is an important observation. Whenever we follow Bender
and Boettcher \cite{BB} and turn our attention to a quasi-Hermitian
quantum model we are forced to endow the given, {\em single} vector
space ${\cal V}$ with {\em two} dif\/ferent inner products. Some
authors \cite{Ali} characterize the second, modif\/ied inner product
by its double bracketing, $\bbr \cdot | \cdot \kkt$. The price to be
paid for this rather unfortunate decision is that the consistent use
of the unmodif\/ied Dirac's notation ceases to be possible.

\subsection{Mostafazadeh's~\cite{Ali} single-Hilbert-space
notation conventions}\label{appendixA.2}

It is not too easy to f\/ind an appropriate notation when a given
phenomenological Hamiltonian is quasi-Hermitian, i.e., manifestly
non-Hermitian in the sense of equation~(\ref{neherme}). The main
dif\/f\/iculty is that in order to get the states properly normalized,
one has to move from the initial, naive Hilbert space ${\cal
H}^{\rm (auxiliary)}$ to the physically correct ${\cal H}^{\rm (standard)}$.

In similar situations, one often feels afraid of using the Dirac's
notation or its analogs\footnote{This is a wide-spread opinion,
especially among mathematicians.}. In practice, such a loss of
contact between the Hermitian and quasi-Hermitian observables would
be rather unpleasant. For this reason people often try to preserve
at least part of this notation. Mostafazadeh~\cite{Ali} of\/fers one
of the latter, modif\/ied notation conventions which became rather
popular in the related literature. Let us brief\/ly recollect some of
its principles and rules, therefore.

First of all we have to imagine that the violation of the
Hermiticity of a given Hamiltonian $H$ in ${\cal H}^{\rm (auxiliary)}$
implies that the eigenstates of $H$ and of its conjugate $H^\dagger$
(as def\/ined in ${\cal H}^{\rm (auxiliary)}$) will be dif\/ferent in
general. This observation has led to the recipe of~\cite{Ali}
where one constructs the two independent series of the respective
eigenstates\footnote{I.e., two series of ket-vector elements of the
{\em same} vector space ${\cal V}_{{\cal H}^{\rm (auxiliary)}}$.}
pertaining to {\em the same} energy eigen\-va\-lues~$E_n$. These states
(numbered by $n = 0,1, \ldots$) must be denoted by two {\em
dif\/ferent} symbols (say, by the ``reserved'' respective Greek letters
$\Phi_n$ and $\Psi_n$) when interpreted as ket- and bra-vectors in
${\cal H}^{\rm (standard)}$,
 \[
 \br \Phi|=\left (|\Phi\kt\right )^\dagger
  \in \big [{\cal
 H}^{\rm (auxiliary)}\big ]' ,\qquad
 \br \Psi|:=\br
 \Phi|\Theta =\left (|\Phi\kt\right )^\ddagger
  \in \big [ {\cal
 H}^{\rm (standard)}\big ]' .
 \label{nedeherme}
 \]
Although we are not going to use such a convention here, we
summarize it, for our readers, in Table~\ref{tabone}. We should
emphasize that as long as we do not stay within the single Hilbert
space ${\cal H}$ of states, two {\em different} def\/initions of the
operation of the Hermitian conjugation emerged in our
considerations.

\begin{table}[h]\centering
\caption{1HS notation of~\cite{Ali} which is {\em not} to be
used here.} \label{pexsp4}
\vspace{1mm}

\begin{tabular}{||c||c|c|c||c||}
\hline \hline
\tsep{1ex}\bsep{1ex} {\rm inner product} &{\rm elements of } ${\cal V}$&{\rm their  duals}
  & {\rm their norms}&  {\rm  Hamiltonians}\ $ H$
 \\
 \hline \hline
\tsep{1ex}\bsep{1ex}  $\br \cdot | \cdot \kt$ (in ${\cal H}^{\rm (auxiliary)}$) & $|\Phi\kt$ & $\br \Phi|=
 (|\Phi\kt )^\dagger
 $ &
 $ \br \Phi|\Phi \kt$ & ``pseudo-Hermitian''\\  \hline
\tsep{1ex}\bsep{1ex}  $\bbr \cdot | \cdot \kkt$ (in ${\cal H}^{\rm (standard)}$) & $|\Phi\kt$  & $
 \br \Psi|=
 (|\Phi\kt )^\ddagger
 $ & $\br \Psi|\Phi \kt$ &
 ``quasi-Hermitian''
 \\ \hline\hline
\end{tabular}
 \label{tabone}
\end{table}

We see that even without switching to the full-f\/ledged 2HS or 3HS
language\footnote{I.e., not speaking openly about the two or three
{\em different} Hilbert spaces ${\cal H}$.}, one must keep trace of
the change of the inner product via an appropriate modif\/ication of
the notation. In this manner one formally re-establishes the
rigorous meaning of the necessary Hermiticity of the Hamiltonian in
the updated, second, physical Hilbert space,
 \[
 H = H^\ddagger\qquad {\rm in }\quad  {\cal H}^{\rm (standard)} .
 \]
As we already mentioned, the so called metric operator (denoted by
symbol $\Theta$ here) can be also used in order to make the
underlying def\/initions less implicit,
 \[
 {\cal O}^\ddagger  \equiv \Theta^{-1}
 {\cal O}^\dagger \Theta .
 \]
In this manner one establishes the new, generalized,
$\Theta$-dependent Hermitian conjugation operation which applies to
any operator ${\cal O}$ acting in ${\cal H}^{\rm (standard)}$.

\section[A few remarks on the simplified Dirac notation
in both the 2HS and 3HS approaches]{A few remarks on the simplif\/ied Dirac notation\\
in both the 2HS and 3HS approaches}\label{appendixB}

As long as the dual-vector def\/inition is inner-product dependent,
the very operation of the (generalized) Hermitian conjugation is
metric-dependent. We recommend that the single-cross superscript
$^\dagger$ stays reserved for its use in ${\cal H}^{\rm (auxiliary)}$
where $\Theta = I$ and that it becomes  paralleled by its
double-cross analogue $^\ddagger$ in ${\cal H}^{\rm (standard)}$ where
$\Theta \neq I$.

\subsection[Metric operator $\Theta$ and the 2HS language]{Metric operator $\boldsymbol{\Theta}$ and the 2HS language}\label{appendixB.1}

A change of the inner product does not necessarily require a {\em
simultaneous} change $\br a|b\kt\to \bbr a|b\kkt$ of {\em both} the
Dirac's bra and ket graphical symbols. This point of view will be
advocated in what follows. Although, formally, its formulation could
start from the idea of selfduality and be made equally rigorous as
in the approach presented in the preceding two subsections, we shall
skip the details here.

Some of the practical merits of such an approach have been discussed
and advocated, e.g., in the study~\cite{knots} of quantum knots. An
even less formal notation convention has been shown working in one
of the oldest papers on the subject of quasi-Hermiticity
\cite{Geyer}. There one f\/inds no double brackets and only the so
called metric operator $\Theta$ appears there\footnote{The original
symbol $T$ is ``translated in Greek'' here since it would interfere
with the time-reversal symbol.} as a key to the transition~(\ref{expo}) between Hilbert spaces.

Even in the 1HS language, the temporary return to Table~\ref{tabone}
and the update $\bbr a|b\kkt \longrightarrow  \br a|\Theta|b\kt$
of the graphical representation of the lower inner product seems
fairly well suited for emphasizing  the inner-product dependence of
the canonical linear functionals. Although such a~1HS convention
seems insuf\/f\/icient in more complicated models, its 2HS amendment of~\cite{timedeprd} proved already sophisticated enough to clarify
the emergence of $\Theta \neq I$ and to render it interpreted as an
introduction of a {\em new} dual space ${\cal V}'_{\text{(non-Dirac)}}$
with {\em modified} elements (i.e., functionals)
 \[
 \mu_b^{\text{(non-Dirac)}}(\cdot) =\br \Theta b| \cdot \kt .
 \]
This version of the 2HS approach can be complemented by the proposal
of a more compact notation as presented, say, in~\cite{knots}.
In essence, one just abbreviates $\mu_b^{\text{(non-Dirac)}}(\cdot)
 \longrightarrow \bbr b| \cdot \kt$.

\subsection{A few details of the transition to the 3HS picture}\label{appendixB.2}

Let us assume that the operators of observables $\Lambda_0,
 \Lambda_1 ,  \ldots$ are admitted non-Hermitian and that, via a
certain non-unitary map $\Omega$, the naive, ill-chosen Hilbert
space ${\cal H}^{\rm (auxiliary)}$ is replaced by another, physical
Hilbert space ${\cal H}^{\rm (standard)}$ where the usual probabilistic
interpretation of the system in question is being restored.
At this point, we shall require that each one of the two roles of
${\cal H}^{\rm (auxiliary)}$ will be carried by a separate Hilbert
space. The f\/irst incarnation of  ${\cal H}^{\rm (auxiliary)}$ will be
denoted by the symbol ${\cal H}^{(F)}$. It will keep playing the
auxiliary role of a mathematically friendly space without any
immediate physical contents. The second, unitarily nonequivalent
space will be treated as the truly physical Hilbert space  ${\cal
H}^{(T)}$ which is, by assumption, neither mathematically easily
accessible nor user-friendly in the context of physics. Its only
comparative advantage will be assumed to lie in a return to the
simplicity of the metric, $\Theta^{(T)}=I$.

After such an absolutely minimal extension of the current Dirac
conventions we may now demand that the third Hilbert space ${\cal
H}^{(T)}$ is built over a {\em new}, independent vector space ${\cal
W} \neq {\cal V}$ with the elements $a$ to be marked by the
specif\/ic, say, spiked version of the Dirac's kets, $a \to |a\pkt$.
The dual space forms again a vector space ${\cal W}'$ with elements
$b = \pbr b|$. As long as, by def\/inition, Hilbert spaces are
self-dual the two vector spaces ${\cal W}$ and ${\cal W}'$ must be
isomorphic\footnote{Without any loss of generality we may work again
with the canonical form of the isomorphism denoted, say, by the
symbol ${\cal T}^{(T)}$ and meaning, in the Dirac's notation, that
${\cal T}^{(T)} : |\psi \pkt \to \pbr \psi|$.}.

\subsubsection{Marking the coexistence of two conjugations}\label{appendixB.2.1}

The two Hilbert spaces ${\cal H}^{(F,S)}$ were def\/ined here over the
same vector space of  kets $|\psi\kt$. In the most common
applications this means that all of these kets may be treated as
linear superpositions of the right eigenkets $|n\kt$ of the {\em
same} upper-case Hamiltonian $H$. As mentioned above, the dif\/ference
between ${\cal H}^{(F)}$ and ${\cal H}^{(S)}$  only emerges during
the introduction of the dual space of functionals. In the former
case the elements of the dual space are def\/ined by the ``standard''
(i.e., Dirac's) Hermitian conjugation. In the language of the so
called ${\cal PT}$-symmetric quantum mechanics \cite{Carl} this
def\/inition is provided by an antilinear operator ${\cal T}={\cal
T}^{(F)}$ from the vector space  ${\cal H}^{(F)}$ in its dual space
$\left ({\cal H}^{(F)}\right )'$. It transforms each ket-vector into
the ``usual'' bra-vector,
 \begin{gather}
 {\cal T}^{(F)}: |\psi\kt \longrightarrow \br \psi| ,
 \qquad
 |\psi\kt \in {\cal H}^{(F)} .
 \label{olnewad}
 \end{gather}
Inside the second Hilbert space ${\cal H}^{(S)}$, in contrast, {\em
another} antilinear operation ${\cal T}={\cal T}^{(S)}$ def\/ines the
{\em different} Hermitian-conjugation operation. At this point one
must pay an enhanced attention to the notation, underlying the
necessity of a {\em clear} distinction between the ``old'' (i.e.,
current, Dirac's) Hermitian conjugation (\ref{olnewad}) employed
inside the f\/irst Hilbert space ${\cal H}^{(F)}$ and the entirely
new, very nonstandard conjugation ${\cal T}^{(S)}$ which is
activated inside the second Hilbert space,
 \begin{gather}
 {\cal T}^{(S)}: \ \ |\psi\kt \longrightarrow \br \br \psi| ,
 \qquad
 |\psi\kt \in {\cal H}^{(S)} .
 \label{newad}
 \end{gather}
In the other words, one is {strongly} recommended to consult a
standard textbook \cite{Messiah} and treat ``brabras'' $ \br \br
\psi|$ literally as {\em linear functionals} in ${\cal H}^{(S)}$.

An easy guide to the acceptance of the alternative, non-Dirac
conjugation (\ref{newad}) appears when we notice that in contrast to
the Dirac's inner product $\br a | b \kt$ between $|a\kt \in {\cal
H}^{(F)}$ and $|b\kt \in {\cal H}^{(F)}$), the new Hilbert space
${\cal H}^{(S)}$ simply allows us to use the new, perceivably less
usual, non-Dirac inner product $\br a |\Theta| b \kt$ between its
elements $|a\kt \in {\cal H}^{(S)}$ and $|b\kt \in {\cal H}^{(S)}$.
Of course, in our present notation we have
 \[
  \br \br \psi_1 | \psi_2 \kt \equiv
  \br \psi_1 |\Theta| \psi_2 \kt .
 \]
This means that while working in  ${\cal H}^{(S)}$ the explicit
remarks concerning the nontrivial metric ${\Theta}$ are redundant.
They must only be made when, for some reason, we must or wish to
turn our explicit attention to the former Hilbert space ${\cal
H}^{(F)}$.

\subsubsection{Simplifying spectral formulae }\label{appendixB.2.2}

Besides the most usual spectral formula representing, say, the
Hamiltonian in ${\cal H}^{(T)}$,
 \[
 \mathfrak{h}=\sum_{n=0}^\infty
 | n
 \pkt E_n \pbr n
|
 \label{speher}
 \]
a perceivably more complicated relation emerges for its  isospectral
partner which is an operator in ${\cal H}^{(F)}$ or ${\cal
H}^{(S)}$,
 \begin{gather}
  H
 =
 \sum_{n=0}^\infty \Omega^{-1}_{}
 | n
 \pkt E_n \pbr n
 | \Omega_{} .
 \label{nespher}
 \end{gather}
Together with the natural innovation of the basis kets in ${\cal
H}^{(F,S)}$,
 \[
  | n
 \kt :=\Omega^{-1}_{}
 | n
 \pkt
 \]
we may also {introduce} another set of ``ketket'' vectors in both
these spaces,
 \[
  | n
 \kt \kt :=
 \Omega^\dagger_{}
 | n
 \pkt \equiv
 \Omega^\dagger_{} \Omega_{}
 | n
 \kt \equiv \Theta
 | n
 \kt
 \
 \in  {\cal H}^{(F,S)} .
 \]
Formula~(\ref{nespher}) then acquires a particularly compact form in
${\cal H}^{(S)}$,
 \[
  H
 =
 \sum_{n=0}^\infty
 | n
 \kt E_n \br \br n
  |
 \]
{\em and} an explicitly metric-dependent form in ${\cal H}^{(F)}$,
 \[
  H
 =
 \sum_{n=0}^\infty
 | n
 \kt E_n \br n
 | \Theta .
 \]
In the same spirit we may deduce that
 \[
 \pbr m|n\pkt = \delta_{m,n}=
 \br \br m | n \kt=
 \br m |\Theta| n \kt ,
 \qquad m,n=0,1,\ldots .
 \]
This means that whenever these sets are perceived as bases, they
should be called {\em orthogonal} in~${\cal H}^{(T,S)}$ and {\em
biorthogonal} in ${\cal H}^{(F)}$.

\subsubsection{Making use of the maps between spaces }\label{appendixB.2.3}

Let us now change and broaden the perspective used during our
discussion of the illustrative example in Section~\ref{sIV} and
assume that we {\em start} from the {knowledge} of the lower-case
Hamiltonian~$\mathfrak{h}$ acting in physical ${\cal H}^{(T)}$ {\em
and} from the {choice} of an {\em arbitrary} orthogonal basis in
${\cal H}^{(S)}$ composed of its brabras $\br \br m|$ and kets
$|m\kt$ which are connected by conjugation~(\ref{newad}). Without
any loss of generality we may further use some suitable unitary
transformations and represent our mapping operator in a simplif\/ied
single-series form
  \[
 \Omega_{}=
 \sum_{n=0}^\infty | n\pkt \mu_{n} \br \br n| .
 \]
{\em All} $\mu_n\in \C\setminus \{0\}$ are {\em independent} free
parameters. By insertion we may immediately verify that these
variable parameters enter also the metric operator,
 \begin{gather*}
 \Theta=\Omega^\dagger \Omega \equiv
 \sum_{n=0}^\infty
 |n\kt \kt \mu_n^* \mu_n \br \br n| .
 \end{gather*}
This operator \cite{Geyer,SIGMA} must be Hermitian, positive
def\/inite and invertible, i.e., the operators
  \[
 \Omega_{}^{-1}=
 \sum_{n=0}^\infty | n\kt \mu_{n}^{-1} \pbr n| ,
 \qquad
 \Theta^{-1}=
 \sum_{n=0}^\infty
 |n\kt \frac{1}{\mu_n^* \mu_n} \br n|
 \]
must be assumed to exist. Under this assumption one could have an
impression that, in principle at least, we might be able to get rid
of the use of the rather exotic concepts of the non-Hermitian
Hamiltonians and/or of the biorthogonal bases in ${\cal H}^{(F)}$.
In practice, this impression proves wrong. In the majority of
applications as reviewed in \cite{Carl} one virtually always works
{\em solely} inside~${\cal H}^{(F)}$ employing both the other two
spaces ${\cal H}^{(T,S)}$ as purely auxiliary theoretical
constructs. Nevertheless, one should keep in mind that whenever we
try to complement the results of our calculations by their correct
probabilistic interpretation, the {\em explicit} use of the two-step
mapping
 \[
 {\cal H}^{(F)} \to {\cal H}^{(S)} \to {\cal H}^{(T)}
 \]
proves unavoidable.

\subsection*{Acknowledgements}

In various stages of development the work has been supported by
Institutional Research Plan AV0Z10480505, by the M\v{S}MT
``Doppler Institute'' project Nr. LC06002, by GA\v{C}R, grant Nr.
202/07/1307 and by the hospitality of Universidad de Santiago de
Chile. Last but not least, three anonymous referees should be
acknowledged for their constructive commentaries.

\pdfbookmark[1]{References}{ref}
\LastPageEnding

\end{document}